\documentstyle[aps,prd]{revtex}
\renewcommand{\baselinestretch}{1.2}
\begin{document}
\large
\title{Elementare Informationsstrukturen der Teilchenphysik\\
       und ihre Verbindung zu Quantenmechanik und Raum-Zeit
       \footnote{Beitrag zur Fr\"uhjahrstagung 2005
       der Deutschen Physikalischen Gesellschaft, Berlin 4. - 9. M\"arz 2005}}
\author{Walter Smilga}
\address{Isardamm 135 d, D-82538 Geretsried}
\address{e-mail: wsmilga@compuserve.com}
\maketitle
\renewcommand{\baselinestretch}{1.2}

\begin{abstract}
Die Bohrsche Feststellung ``Physikalische Ph\"anomene werden
\begin{em}relativ\end{em} zu \begin{em}verschiedenen\end{em} experimentellen
Anordnungen beobachtet" wird angewandt auf ein System von bin\"aren Elementen
als Tr\"ager elementarer Informationseinheiten.
Im Sinne von Bohr wird eine Beschreibung \begin{em}relativ\end{em} zu
``makroskopischen" Anordnungen solcher Elemente formuliert.
Diese erfordert die Einf\"uhrung eines Hilbertraum-Formalismus.
Es wird gezeigt, da{\ss} der Hilbertraum symmetrisch bez\"uglich der
de~Sitter-Gruppe SO(3,2) ist.
F\"ur makroskopische Anordnungen wird letztere durch die Poincar\'e-Gruppe
approximiert.
Man erh\"alt dadurch ein relativistisches Raum-Zeit-Kontinuum als Ausdruck
der Orientierung makroskopischer Anordnungen \begin{em}relativ\end{em}
zueinander.
Einzelnen Bin\"arelementen l\"a{\ss}t sich dann ebenfalls eine ``Position"
\begin{em}relativ\end{em} zu makroskopischen Bezugselementen zuordnen.
Damit erscheinen die Bin\"arelemente dem Beobachter als massive
Spin-1/2-Teilchen.

Der informationstheoretische Zugang bestimmt eine Massenskala, liefert
stringente Ans\"atze f\"ur alle vier Wechselwirkungen und legt prinzipiell
Kopplungskonstanten und Massen fest.
Er stellt damit trotz seiner Einfachheit eine potentielle Basis f\"ur eine
Teilchentheorie dar, die \"uber das Standardmodell hinausf\"uhrt.
\end{abstract}


\renewcommand{\baselinestretch}{1.0}

\section{Einleitung}

In den Entstehungsjahren der Quantenmechanik fand eine intensive Diskussion
statt \"uber die Bedeutung dieser neuen Theorie und \"uber die Art und Weise,
wie wir im atomaren Bereich zu Erkenntnissen gelangen.
Diese Diskussion f\"uhrte zur Kopenhagener Deutung, die ein, wenn
auch nicht endg\"ultig befriedigendes, so doch f\"ur die damalige Zeit
ausreichendes Verst\"andnis der Quantenwelt erm\"oglichte.

In der heutigen Teilchenphysik vermi{\ss}t man eine entsprechende
Grundlagendiskussion.
Innerhalb des Standardmodells hat die suggestive bildliche Kraft der
Feynman-Graphen zu einem Scheinverst\"andnis gef\"uhrt, das einzelne
Beitr\"age zur st\"orungstheoretischen Entwicklung als direkte Abbilder
physikalischer Elementarprozesse interpretiert.
Dies hat zu teilweise recht phantasievollen Deutungen der
Elementarteilchenphysik Anla{\ss} gegeben, war aber einer sachbezogenen
Diskussion nicht eben f\"orderlich.
Die neueren theoretischen Ans\"atze sind im Bereich der Planck-L\"ange
angesiedelt, der der Empirie und der sich daran orientierenden physikali\-schen
Begriffswelt grunds\"atzlich entzogen ist.

Unter diesen Umst\"anden scheint es geboten, sich \"uber solche Grundlagen 
unserer Disziplin Klarheit zu verschaffen, die der Diskussion zug\"anglich 
sind und von denen wir annehmen d\"urfen, da{\ss} sie von allgemeiner
G\"ultigkeit und nicht Produkte freier Interpretationen sind.

Ich werde dazu zwei Aussagen aus der Fr\"uhzeit der Quantenmechanik
aufgreifen, die uns heute mehr oder weniger selbstverst\"andlich erscheinen,
und zeigen, da{\ss} sich aus dem scheinbar Selbst\-verst\"andlichen
grundlegende Einsichten in die Struktur einer realistischen Theorie der
Elementarteilchen gewinnen lassen.

Von Niels Bohr stammt folgende Feststellung \cite{mj}:
\begin{quote}
``Physika\-lische Ph\"anomene werden \textit{relativ} zu
\textit{verschiedenen} experimentellen Anordnungen beobachtet."
\end{quote}
F\"ur Bohr lag die Betonung offensichtlich auf dem Wort ``verschieden".
Er wollte zum Ausdruck bringen, da{\ss} sich innerhalb der Quantenmechanik
etwa Elektronen je nach experimenteller Anordnung ent\-weder wie Teilchen
oder wie eine Welle verhalten.
In meinem Beitrag wird die Betonung mehr auf ``re\-lativ" liegen, - nicht so
sehr, weil ich mich dem Einstein-Jahr 2005 verpflichtet f\"uhle, sondern, weil
dieses Wort in Bohrs Satz, wie wir sehen werden, einen Schl\"ussel liefert zu
einem tieferen Verst\"andnis sowohl der Quantenmechanik als auch von
Raum-Zeit.
Mit Verst\"andnis meine ich: es erlaubt Quantenmechanik und Raum-Zeit als
Konsequenz allgemeiner logischer Gesetzm\"a{\ss}igkeiten zu verstehen.

Ich lege meinem Beitrag ein weiteres Zitat von Bohr \cite{bh} zugrunde:
\begin{quote}
``Es ist falsch anzunehmen, es sei Aufgabe der Physik herauszufinden, wie
die Natur \begin{em}ist\end{em}.
Physik betrifft das, was wir \"uber die Natur \begin{em}sagen\end{em}
k\"onnen."
\end{quote}
Physik ist demnach eine Angelegenheit der Gewinnung, Aufbereitung und
Verarbeitung von Information \"uber die Natur.
Teilchenphysik sucht dementsprechend nach den ``kleinsten" Bausteinen
innerhalb unseres Wissens \"uber die Natur.
Unter den kleinsten Bausteinen m\"ussen wir in diesem Zusammenhang offenbar
solche Informationsstrukturen verstehen, die keine weitere Unterteilung
zulassen.
Lassen Sie mich dies als Arbeitshypothese so formulieren:\\
``Die Teilchenphysik sucht, im physikalischen Kontext, nach den
kleinsten Bausteinen von Information."

Nun, die kleinsten Bausteine von Information sind wohlbekannt:
es sind dies die Bits (binary digits) der Informatik.
Welche Eigenschaften kommen dann aber diesen Bits innerhalb der
Teilchenphysik zu?
K\"onnen wir Bits im Experiment beobachten?
Gibt es eine Verbindung zwischen Bits und den Begriffen, mit denen wir
normalerweise hantieren, wenn wir Physik betreiben, als da sind:
Raum-Zeit, Symmetrien, Teilchen und Wechselwirkung?

Ich werde in meinem Beitrag eine solche Verbindung herstellen.
Und ich will Ihnen zeigen, da{\ss} einige Ph\"anomene, die Sie bisher
f\"ur grundlegende Eigenschaften der Natur hielten, tats\"achlich als
nat\"urliche Eigenschaften grundlegender Informationsstrukturen
verstanden werden k\"onnen.

Es hat seit mehr als 30 Jahren Versuche gegeben, mittels elementarer
informationstheoreti\-scher Ans\"atze grundlegende Fragen der Physik zu
beantworten.
Ich rechne hierzu Roger~Penrose's\cite{rp} Versuch aus den 70er-Jahren,
aus einem Netzwerk von zweikomponentigen Spinoren Eigenschaften von
Raum-Zeit abzuleiten.
Er wurde damit zum Begr\"under der immer noch hochaktuellen
Forschungsrichtung der Spin-Netzwerke.
Ein \"Uberblick hierzu findet sich in \cite{jcb}.
Die Arbeiten an Spin-Netzwerken verfolgen nach wie vor das Ziel, Raum-Zeit
aus grundlegenden Strukturen herzuleiten und suchen dabei, wie auch die
String-Theorien, eine Antwort im Bereich der Planck-L\"ange.
Dies f\"uhrt allerdings zur bisher ungel\"osten Frage:
Wie kommt man von der Planck-L\"ange zu physikalisch zug\"anglichen
Dimensionen?

Etwa zur gleichen Zeit wie Penrose hat sich Carl Friedrich von Weizs\"acker
\cite{cfw} in seinem Starnberger Institut mit informationstheoretischen
Ans\"atzen besch\"aftigt.
Letztere waren sehr allgemein gefa{\ss}t und widersetzten sich daher
Versuchen, eine konkrete Verbindung zur Empirie herzustellen.
Eine moderne Variante von Weizs\"ackers ``bin\"aren Alternativen" finden
wir heute in der Quanteninformatik unter der Bezeichnung \textit{Q-bit}.

Beiden Ans\"atzen, sowohl Penrose's als auch von Weizs\"ackers ist gemeinsam,
da{\ss} sie die quantenmechanische Beschreibung als selbstverst\"andlichen
Teil ihres Ansatzes betrachteten, der nicht weiter zu hinterfragen ist.
Ich werde Ihnen zeigen, da{\ss} die a-priori-Annahme einer Quantenmechanik
nicht nur unn\"otig ist, sondern zudem den Blick auf tiefere
Zusammenh\"ange verstellt.
Die quantenmechanische Beschreibung wird sich dann zwangsl\"aufig als
Konsequenz der ersten Bohrschen Feststellung ergeben, und zwar in einer
transparenten Weise, die keine weiteren Interpretationsprobleme aufwirft.

\section{Von bin\"aren Elementen zu Spinoren}

Beginnen wir also ganz elementar mit einer Menge $B$ von bin\"aren
Elementen $z$, die genau zwei Zust\"ande annehmen k\"onnen.
Den Gepflogenheiten der Informatik folgend, bezeichne ich diese 
mit $0$ und $1$:
\begin{equation}
B = \{ z \, | \, z \in \{0, 1\} \} .                              \label{2-0}
\end{equation}

Die Elemente $z$ werden allein durch ihre Zust\"ande $0$ und $1$
unterschieden, sind aber im \"ubrigen nicht-unterscheidbar,
denn w\"aren sie es, m\"u{\ss}ten sie noch weitere Merkmale aufweisen, was
aber per definitionen nicht der Fall ist.

Diese Elemente sollen dazu verwendet werden, darin Information abzulegen,
\"ahnlich den Bits innerhalb der Arbeitsspeichers eines Rechners.
Das Ablegen bin\"arer Information mit den m\"oglichen Auspr\"agungen $a$
und $b$ in ein Bit ist ein Isomorphismus den Menge $\{a, b\}$ in die Menge
$\{0, 1\}$.
Es gibt genau zwei derartige Isomorphismen, n\"amlich
\begin{equation}
  a \rightarrow 0, \, \, b \rightarrow 1                      \label{2-a}
\end{equation}
und
\begin{equation}
  a \rightarrow 1, \, \, b \rightarrow 0 \,.                  \label{2-b}
\end{equation}
In der Informatik w\"urde man von zwei unterschiedlichen Kodierungen sprechen.
Da wir nicht vor\-aussetzen k\"onnen, da{\ss} die Natur nur einen der
beiden Isomorphismen realisiert (wenn doch, welchen, und wie
w\"aren sie zu unterscheiden?), werden wir beide M\"oglichkeiten
ber\"ucksichtigen m\"ussen.

Wir werden im folgenden Untermengen von $B$ begegnen, die wir dazu verwenden,
``makroskopische Anordnungen" zu bilden. Unter makroskopischen Anordnungen
will ich Teilmengen von, sagen wir, etwa $10^{23}$ Elementen verstehen.
Eine makroskopische Anordnung stellt eine Informationsmenge dar, die
beispielsweise geeignet ist, die Gesamtinformation \"uber einen
makroskopischen physikali\-schen K\"orper aufzunehmen.
Die Gesamtmenge $B$ sollte deutlich gr\"o{\ss}er sein, und, sagen wir, etwa
$10^{80}$ Elemente umfassen.

Mit diesen Elementen werde ich versuchen, Physik zu treiben.
Was meine ich mit Physik in diesem Zusammenhang?
Denken wir an Bohr: Physikalische Ph\"anomene werden beobachtet relativ zu
verschiedenen experimentellen Anordnungen.

Also, eine experimentelle Anordnung wird sich, zumindest im
Gedankenexperiment, durch die erw\"ahnte makroskopische Anordnung
beschreiben lassen.
Und ein physikalisches Ph\"anomen ist wohl im einfachsten Fall ein
weiteres, einzelnes bin\"ares Element.
Was ist dann unter einem bin\"aren Element \begin{em}relativ\end{em} zu
verschiedenen makroskopischen Anordnung zu verstehen?
Offensichtlich f\"uhrt uns diese Frage auf dem Begriff der
\textit{Relativit\"at von Information.}
Diese Erweiterung des Relativit\"atsbegriff auf die einfachste
informationstheoretische Einheit wird sich im folgenden als \"au{\ss}erst
fruchtbar erweisen.

Um mit diesem Begriff praktisch umgehen zu k\"onnen, ben\"otigen wir
zun\"achst eine geeignete formale Sprache.
Wir m\"ussen uns also ein mathematisches Begriffssystem zurechtlegen,
das sich aus den elementaren Eigenschaften bin\"arer Elemente ableitet.

Ich beginne, indem ich bin\"are Elemente durch bin\"are \textit{Zust\"ande}
in Vektorform darstelle, also
\begin{equation}
|1\rangle :=  \left( \begin{array}{*{1}{c}} 1 \\ 0 \end{array} \right)
 \;\; \mbox{ und } \;\;
|0\rangle := \left( \begin{array}{*{1}{c}} 0 \\ 1 \end{array} \right) .
                                                                \label{2-1}
\end{equation}
Dann f\"uhre ich die Matrix
\begin{equation}
\sigma_3 := \left( \begin{array}{*{2}{c}} 1 & 0 \\ 0 & -1 \end{array} \right)
                                                                \label{2-2}
\end{equation}
ein. Mit ihr k\"onnen wir z.B. eine einfache Eigenwert-Gleichung aufstellen
\begin{equation}
\sigma_3 |z\rangle = \zeta |z\rangle ,                          \label{2-3}
\end{equation}
wo $|z\rangle$ entweder $|1\rangle$ oder $|0\rangle$, und $\zeta$
die zugeh\"origen Eigenwerte $1$ bzw. $-1$ sind.
Ich will auch gleich einen \begin{em}Erwartungswert\end{em} von $\sigma_3$
durch
\begin{equation}
\langle z| \sigma_3 |z\rangle = \zeta ,                         \label{2-4}
\end{equation}
definieren, wobei $\langle z|$ f\"ur den zu  $|z \rangle$ transponierten
Vektor steht.
Wir k\"onnen auf diese Weise ``mathe\-matisch exakt" feststellen, welche
Ausrichtung das durch $z$ beschriebene bin\"are Element hat.

Eine weitere Matrix
\begin{equation}
\sigma_1 := \left( \begin{array}{*{2}{c}} 0 & 1 \\ 1 & 0 \end{array} \right),
                                                                \label{2-5}
\end{equation}
angewandt auf $|1\rangle$ oder $|0\rangle$, \"andert $|1\rangle$ in
$|0\rangle$ bzw. $|0\rangle$ in $|1\rangle$.
Da unser Formalismus keine Einstellung der bin\"aren Elements bevorzugen soll,
definiert $\sigma_1$ eine diskrete \begin{em}Symmetrietransformation\end{em}
innerhalb der Zust\"ande 0 und 1.

Unsere Information \"uber die Ausrichtung eines bin\"aren Elements mu{\ss}
nicht immer vollst\"andig sein. Das hei{\ss}t, wir m\"ussen damit rechnen,
da{\ss} unsere Information besagt, da{\ss} mit einem gewissen Grad an
Sicherheit der Zustand 1 realisiert ist, aber m\"oglicherweise doch
der Zustand 0 vorliegt.
Ich versuche, diesem Umstand gerecht zu werden, indem ich auch
Linearkombinationen der Form
\begin{equation}
|s\rangle = \alpha \, |1\rangle + \beta \, |0\rangle  ,         \label{2-6}
\end{equation}
betrachte, wo $\alpha$ und $\beta$ komplexe Zahlen sind mit der
Einschr\"ankung
\begin{equation}
\alpha^* \alpha + \beta^* \beta = 1 .                           \label{2-7}
\end{equation}
Ein solcher Mischzustand soll als (zweikomponentiger) \begin{em}Spinor\end{em}
bezeichnet werden.
Mit diesem Schritt habe ich einen \begin{em}komplexen Vektorraum\end{em} mit
$|1\rangle$ und $|0\rangle$ als Basisvektoren eingef\"uhrt.

Die Mischzust\"ande stellen eine Erweiterung des bisherigen Formalismus dar.
Diese Erweiterung ist wohldefiniert. Sie verletzt keine Regel der Logik.
Folglich ist gegen sie grunds\"atzlich nichts einzuwenden.
Sie wird letztlich durch ihren praktischen Nutzen zu rechtfertigen sein.
Vorl\"aufig begr\"unde ich die Mischzust\"ande mit dem Wunsch, auch eine
unvollst\"andige Information \"uber ein bin\"ares Element darstellen
zu k\"onnen.

Ich definiere au{\ss}erdem ein Skalarprodukt zwischen zwei Zust\"anden
$|s_1\rangle$ und $|s_2\rangle$ durch
\begin{equation}
\langle s_1 | s_2 \rangle = \alpha^*_1 \alpha_2 + \beta^*_1 \beta_2  .
                                                                 \label{2-9}
\end{equation}
Damit wird aus den komplexen Vektorraum ein \begin{em}Hilbertraum\end{em}.
Im Sprachgebrauch der Quanten\-informatik wird dieser Hilbertraum als
\begin{em}Q-bit\end{em} bezeichnet.
Mit der Einf\"uhrung von Q-bits erhalten wir Zust\"ande,
die zwischen den Einstellm\"oglichkeiten der klassischen Bits oder
\begin{em}C-bits\end{em} ``interpolieren".

Der Hilbertraum-Formalismus erm\"oglicht einige mathematische
Operationen, die in der rein bin\"aren Beschreibung nicht m\"oglich sind.
Beispielsweise k\"onnen wir die Algebra der Matrizen $\sigma_1$ und
$\sigma_3$ bez\"uglich des Kommutationsproduktes betrachten. Wir
erhalten dann eine dritte Matrix $\sigma_2$ mittels des Kommutators
\begin{equation}
2\,i\,\sigma_2 := [\sigma_3, \sigma_1]                            \label{2-10}
\end{equation}
und zwar in der Form
\begin{equation}
\sigma_2 = \left( \begin{array}{*{2}{c}} 0 & -i \\ i & 0 \end{array} \right).
                                                                 \label{2-11}
\end{equation}

$\sigma_1$, $\sigma_2$ und $\sigma_3$ sind die bekannten Pauli-Matrizen und,
wie Sie wissen, erf\"ullen diese die Vertau\-schungs\-relationen der SU(2).
Sie sind damit die infinitesimalen Erzeugenden von SU(2)-Transformationen
innerhalb des Hilbertraums der zweikomponentigen Spinoren.
Diese Symmetrie-Transformationen sind kontinuierlich im Gegensatz zu der
diskreten Symmetrie-Operation an den Basis-Zust\"anden. Sie sind eine
unmittelbare Konsequenz der diskreten Symmetrie-Eigenschaft der bin\"aren
Zust\"ande, die sich auf diese Weise auf den interpolierenden Hilbertraum
\"ubertr\"agt.

\section{ Begr\"undung der Quantenmechanik}

Es ist Ihnen sicher nicht entgangen, da{\ss} wir soeben eine
quantenmechanische Beschreibung eines bin\"aren Elements eingef\"uhrt
haben.
Wir sind dabei von der Tatsache ausgegangen, da{\ss} die elementaren
Einheiten der Information bereits in diskreter, das hei{\ss}t quantisierter
Form vorliegen.
Unter dem Aspekt unvollst\"andiger Information haben wir
zur mathematischen Beschreibung dieser Informations-Quanten einen
Hilbertraum-Formalismus eingef\"uhrt.
Dieser Formalismus gibt uns nun Begriffe an die Hand, mit denen wir Beziehungen
innerhalb eines Systems von bin\"aren Elementen beschreiben k\"onnen.
Schauen wir also, wie dieser Formalismus praktisch einzusetzen ist.

Wie Ihnen aus der Quantenmechanik der Drehimpulse gel\"aufig ist, kann man
Spinor-Zust\"ande verkoppeln.
Zwei Spinoren ergeben dann (nach Ausreduktion) entweder einen Gesamtdrehimpuls
null oder einen Drehimpuls eins mit drei Einstellm\"oglichkeiten $-1$, $0$ und
$+1$.
Dieses Spiel l\"a{\ss}t sich mit einer gr\"o{\ss}eren Zahl von
Spinoren wiederholen und man kommt dann schlie{\ss}lich zum klassischen
Grenzfall, wo sich dieses quantenmechanische Gebilde verh\"alt wie eine
starrer K\"orper, der in drei Winkelrichtungen frei orientierbar ist.
Auf diese Weise verhilft uns der Hilbertraum-Formalismus zun\"achst einmal
zu einer handlichen Beschreibung makroskopischer Anordnungen von Spinoren,
im folgenden kurz als Makrosysteme bezeichnet.
Entsprechend der vorstehenden Konstruktion ist auch ein Makrosystem seinem
Wesen nach Information, die sich aus elementaren Informationseinheiten
zusammensetzt.

Nehmen wir an, da{\ss} wir ein Makrosystem hergestellt haben, das wir
im Sinne der Bohrschen ``experimentellen Anordnung" verwenden k\"onnen.
Eine ``Verschiedenheit" der Anordnungen k\"onnen wir dadurch realisieren,
da{\ss} wir das Makrosystem in unterschiedlichen Winkel-Orientierungen
einsetzen.
Richten wir nun ein einzelnes bin\"ares Element relativ zu dieser
Anordnung so aus, da{\ss} das Element im Zustand 1 relativ zur Anordnung
erscheint.
Dann drehen wir die Anordnung durch Anwendung einer SU(2)-Transformation
um genau 90 Grad.
Wenn wir dabei das Koordinatensystem mitbewegen, bildet sich diese Drehung
mittels der entsprechenden SU(2)-Transformation auf den Spinor-Zustand ab.
Wir erhalten dann einen Zustandsvektor, der zu gleichen Teilen aus den
Zust\"anden 1 und 0 besteht. Was bedeutet das?

Wenn wir den Versuch unternehmen, die Richtung des Spinors relativ
zur gedrehten Anordnung zu ``messen", d\"urfen wir annehmen, ohne Details
des damit verbundenen Me{\ss}vorgangs zu kennen, da{\ss} das Ergebnis
schon allein aus Symmetriegr\"unden unbestimmt sein wird.
D.h. wir werden, wenn wir das Experiment gen\"ugend h\"aufig ausf\"uhren,
mit 50\% Wahrscheinlichkeit einen Zustand 1 feststellen
und mit ebensolchen 50\% einen Zustand 0.
Eben dieses Ergebnis k\"onnen wir der mathemati\-schen Darstellung des
bin\"aren Elements durch den Hilbertraum-Zustand entnehmen, wenn wir die
Koeffizienten der Basiszust\"ande als Wahrscheinlichkeitsamplituden
interpretieren.

Ein Hilbertraum-Zustand bietet sich also konkret dazu an,
die Information \"uber den Zustand des bin\"aren Elements,
entsprechend der Bohrschen Feststellung, \begin{em}relativ zu einer
experimentellen Anordnung\end{em} zu beschreiben. Diese Beschreibung
wird im allgemeinen die Form einer Wahrscheinlichkeits\-aussage haben.
Und zwar dann, wenn das Ergebnis der Messung mehr oder weniger unbe\-stimmt
ist.
Unbestimmtheit ist eine legitime Auspr\"agung von Information, und
innerhalb der Quantenmechanik ist der Begriff der Unbestimmtheit bekanntlich
von zentraler Bedeutung.

Beachten Sie bitte, da{\ss} der Hilbertraum als beschreibendes Werkzeug
eingef\"uhrt wurde und nicht eine grunds\"atzliche Eigenschaft unserer Menge
von bin\"aren Elementen darstellt.
Um mit Bohr zu sprechen \cite{bh2}:
\begin{quote}
``Es gibt keine Quantenwelt.
Es gibt nur eine abstrakte quantenmechanische Beschreibung."
\end{quote}

Der spezielle quantenmechanische Charakter im Sinne einer
Wahrscheinlichkeitsaussage ist hier
durch die Tatsache bedingt, da{\ss} das Makrosystem mehr
Orientierungsm\"oglichkeiten besitzt als das bin\"are Element.

Die quantenmechanische Beschreibung f\"uhrt im System von
bin\"aren Elementen zu drei wichtigen Ergebnissen:
\begin{itemize}
\item
Zum einen definiert sie, was vielleicht \"uberraschend ist, ``klassische"
Freiheitsgrade zur Beschreibung von Makrosystemen;
in unserem Fall liefert sie die drei Winkelfreiheitsgrade, um die sich ein
makroskopisches Gebilde quasi-kontinuierlich drehen l\"a{\ss}t.
\item
Zum zweiten erlaubt sie eine Beschreibung mikroskopischer Elemente relativ
zu beliebig orientierten makroskopischen Gebilden; hier wird der
``quantenmechanische" Aspekt deutlich.
\item
Und schlie{\ss}lich formuliert sie die Symmetrieeigenschaften des Systems
durch eine Lie-Gruppe: wir erhalten eine Symmetrie bez\"uglich SU(2), 
wenn wir einzelne Spinoren betrachten, und SO(3) in Zusammenhang mit 
Makrosystemen.
\end{itemize}

\section{ Dirac-Spinoren und de~Sitter-Gruppe}

Lassen Sie mich noch einmal auf die Darstellung der Zust\"ande durch
Spaltenvektoren zur\"uck\-kommen.
Erinnern wir uns daran, da{\ss} das Ablegen von Information in ein Bit
grunds\"atzlich zwei verschiedene Isomorphismen / Kodierungen zul\"a{\ss}t.
Die Zuordnung
\begin{equation}
  a \rightarrow |1\rangle, \, \, b \rightarrow |0\rangle      \label{3-1a}
\end{equation}
realisiert einen solchen Isomorphismus.
Der zweite ist dann so anzusetzen:
\begin{equation}
  a \rightarrow |0\rangle, \, \, b \rightarrow |1\rangle  .   \label{3-1b}
\end{equation}
Durch Zusammenfassen beider Alternativen zu vierkomponentigen
Spaltenvektoren bringen wir unseren Formalismus in eine allgemeing\"ultige
Form, die es erlaubt, beide Kodierungen gleichberechtigt zu behandeln:
\begin{eqnarray}
|u_a\rangle &:=& \left(\begin{array}{*{1}{c}} 1\\0\\0\\0 \end{array}\right)
 \; \mbox{ , } \;
|u_b\rangle  :=  \left(\begin{array}{*{1}{c}} 0\\1\\0\\0 \end{array}\right)
\; \mbox{ , } \;                                               \label{3-2a}\\
\nonumber \\
|v_a\rangle &:=& \left(\begin{array}{*{1}{c}} 0\\0\\0\\1 \end{array}\right)
 \; \mbox{ , } \;
|v_b\rangle  :=  \left(\begin{array}{*{1}{c}} 0\\0\\1\\0 \end{array}\right).
                                                               \label{3-2b}
\end{eqnarray}

Auch hier f\"uhre ich einen interpolierenden Vektorraum ein und erhalte dann
vierkomponentige Spinoren, die ich aus naheliegenden Gr\"unden
als Dirac-Spinoren bezeichnen will.
An Stelle der Pauli-Matrizen haben wir jetzt $4\times4$ Spin-Matrizen:
\begin{equation}
\sigma_{ij} := \epsilon_{ijk} \left(\begin{array}{*{2}{c}}
\sigma_k & 0 \\ 0 & \sigma_k \end{array}\right),
\; i,j,k = 1,2,3,                                               \label{3-3}
\end{equation}
wobei $\epsilon_{ijk}$ das Permutationssymbol ist.
Die Matrix
\begin{equation}
\gamma^0 :=
\left( \begin{array}{*{2}{c}} I & 0 \\ 0 & -I \end{array} \right),
                                                                \label{3-4}
\end{equation}
wobei $I$ die $2\times 2$ Einheitsmatrix ist,
liefert den Eigenwert $+1$ bei Anwendung auf die erste Gruppe von
Basis-Spinoren
(\ref{3-2a}), und $-1$ bei Anwendung auf die zweite (\ref{3-2b}).

Analog zu $\sigma_1$ definiere ich die Matrix
\begin{equation}
\tau := \left( \begin{array}{*{2}{c}} 0 & I \\ I & 0 \end{array} \right),
                                                                \label{3-4a}
\end{equation}
die die erste mit der zweiten Gruppe, also die Kodierungen, vertauscht.
Da beide Kodierungen gleich zu behandeln sind, definiert diese
Austauschoperation wieder eine Symmetrie-Transformation innerhalb unseres
Formalismus.
Dann ist auch die kombinierte Anwendung von $\sigma_1$ und $\tau$
eine Symmetrie-Operation.
Sie ist gegeben durch die Matrix
\begin{equation}
\gamma^1 :=
\left( \begin{array}{*{2}{c}} 0 & \sigma_1 \\ \sigma_1 & 0 \end{array}\right).
\label{3-5}
\end{equation}

Wenn wir die Algebra der bisher definierten Matrizen $\gamma^0$, $\gamma^1$
und $\sigma_{ij}$ bez\"uglich des Kommutatorprodukts betrachten, erhalten wir
weitere Matrizen
\begin{equation}
\gamma^k :=
\left( \begin{array}{*{2}{c}} 0 & \sigma_k \\ \sigma_k & 0 \end{array}\right)
                                                                \label{3-6}
\end{equation}
und
\begin{equation}
\sigma^{0k} = -\sigma^{k0} := \left( \begin{array}{*{2}{c}} 0 & i\sigma_k \\
       -i\sigma_k & 0 \end{array}\right), \mbox{ wobei } k=1,2,3.\label{3-6a}
\end{equation}

Die Matrizen (\ref{3-4}) und (\ref{3-6}) sind Dirac's $\gamma$-Matrizen in der
sogenannten Dirac-Darstellung. Sie erf\"ullen die wohlbekannten
Antivertauschungsrelationen
\begin{equation}
\{\gamma_\mu, \gamma_\nu \} = 2 g_{\mu\nu} .                    \label{3-7a}
\end{equation}
sowie diese Vertauschungsrelationen
\begin{equation}
\frac{i}{2} \, [\gamma_\mu, \gamma_\nu ] = \sigma_{\mu\nu},     \label{3-7b}
\end{equation}
wobei jetzt $\mu, \nu = 0,\ldots,3$.

Um zu einem Hilbertraum zu kommen, definiere ich noch ein Skalarprodukt,
wie Sie es aus der Dirac-Theorie des Elektrons kennen,
\begin{equation}
\langle \bar{a} | b \rangle  \mbox{   mit   }
\langle \bar{a} | = \langle a | \gamma^0  .                     \label{3-8}
\end{equation}

Die $4\times4$-Matrizen
\begin{equation}
s_{\mu\nu} :=\, \frac{1}{2} \sigma_{\mu\nu}
\mbox{   und  }
s_{\mu4} :=\, \frac{1}{2} \gamma_\mu ,                          \label{3-9}
\end{equation}
die wir aus den Dirac-Matrizen erhalten,
bilden eine irreduzible Darstellung der \textit{de~Sitter-Gruppe} SO(3,2)
auf dem Hilbertraum der Dirac-Spinoren.
Man beweist dies durch Nachpr\"ufen der Vertauschungsrelationen der SO(3,2):
\begin{equation}
[s_{\mu\nu}, s_{\rho\sigma}] =
-i[g_{\mu\rho} s_{\nu\sigma} - g_{\mu\sigma}
s_{\nu\rho} + g_{\nu\sigma} s_{\mu\rho}
- g_{\nu\rho} s_{\mu\sigma}] \mbox{ , }                         \label{3-10}
\end{equation}
\begin{equation}
[s_{\mu4}, s_{\nu4}] = -i s_{\mu\nu} \mbox{ , }                 \label{3-11}
\end{equation}
\begin{equation}
[s_{\mu\nu}, s_{\rho4}]
= i[g_{\nu\rho} s_{\mu4} - g_{\mu\rho} s_{\nu4}]  .             \label{3-12}
\end{equation}

Die SO(3,2)-Transformationen sind eine unmittelbare Konsequenz der diskreten
Symmetrieope\-rationen zwischen den vierkomponentigen Basiszust\"anden. Sie
erweitern diese Operationen auf den interpolierenden Hilbertraum der
Dirac-Spinoren.
Eine Untergruppe der SO(3,2) ist die homogene Lorentz-Gruppe SO(3,1) mit den
Vertauschungsrelationen (\ref{3-10}).
Die gemeinsame Untergruppe SO(3) entspricht der SU(2)-Symmetrie der
zweikomponentigen Spinor-Anteile des Dirac-Spinors.

Wir erhalten damit die de~Sitter-Gruppe SO(3,2) als die nat\"urliche
Symmetriegruppe einer Informationsstruktur, die auf bin\"aren Elementen
basiert.
Eingegangen ist dabei die Art der Beobachtung eines bin\"aren
Elements relativ zu einer makroskopischen Anordnung, was uns zur
Hilbertraum-Beschreibung gef\"uhrt hat, sowie die Gleichbehandlung der
beiden Einstellungen des bin\"aren Elements, also der Spin-Richtung, und
die Gleichbehandlung der beiden Kodierungsm\"oglichkeiten.

\section{Poincar\'{e}-Gruppe und Raum-Zeit}

Es ist jetzt nur noch ein kleiner Schritt zur Poincar\'e-Gruppe.
Betrachten wir wieder eine makroskopische Anordnung und dessen
Zust\"ande mit gro{\ss}en quasi-kontinuierlichen Quantenzahlen.
Nehmen wir an, wir haben die Zust\"ande so ausgerichtet, da{\ss} die
Erwartungswerte der Operatoren $S_{\mu 4}$ gro{\ss} sind im Vergleich zu
denen von $S_{\mu\nu}$. 
(Die Operatoren $S$ sind definiert als Summe der Operatoren $s$ der 
einzelnen Spinoren.
Ich verwende Gro{\ss}buchstaben f\"ur Operatoren und Zust\"ande von 
Makrosystemen.)

Dann k\"onnen wir die bekannte Methode der Gruppenkontraktion \cite{iw,fg}
auf diesen Teil des Hilbert\-raums anwenden und erhalten die
Poincar\'e-Gruppe P(3,1) als approximative Symmetriegruppe.
Sie k\"onnen dies leicht nachvollziehen, wenn Sie den Kommutator (\ref{3-11})
unter den genannten Bedingungen betrachten.
Wenn $S_{\mu 4}$ gro{\ss} ist gegen\"uber $S_{\mu\nu}$, kann der Kommutator
approximiert werden durch
\begin{equation}
[P_\mu, P_\nu] = 0,                                             \label{4-2}
\end{equation}
worin die Operatoren $P_\mu$ die Approximation von $S_{\mu 4}$ darstellen.
Aus (\ref{3-12}) leiten sich dann die Vertauschungsrelation der $P_\mu$
mit den Erzeugenden der Lorentz-Transformationen ab
\begin{equation}
[S_{\mu\nu}, P_\rho]
= i[g_{\nu\rho} P_\mu - g_{\mu\rho} P_\nu]  .                   \label{4-3}
\end{equation}
(\ref{3-10}), (\ref{4-2}) und (\ref{4-3}) bilden zusammen die
Vertauschungsrelationen der Poincar\'e-Gruppe.
An die Stelle des quasi-kontinuierlichen Spektrums der SO(3,2)-Quantenzahlen
tritt jetzt das kontinuierliche Spektrum der $P_\mu$.

Die Eigenzust\"ande der $P_\mu$ k\"onnen wir verwenden, um in bekannter Weise
neue Zust\"ande $|X\rangle$ zu definieren, die in raumartigen Richtungen
lokalisiert sind,
\begin{equation}
|X\rangle := (2\pi)^{-3/2} \int\limits^\infty_{-\infty}\! d^3P
\: e^{i x^\mu P_\mu}\, |P\rangle.                                \label{4-4}
\end{equation}
Die Operatoren $P_\mu$ erzeugen dann bei Anwendung auf diese Zust\"ande
Translationen mit Verschiebungsvektoren $a$
\begin{equation}
e^{ia^\mu P_\mu} \, |X\rangle = |X + a\rangle .                  \label{4-5}
\end{equation}
Die Verschiebungsvektoren spannen dabei ein 4-dimensionales Raum-Zeit-Kontinuum
mit Minkowski-\-Metrik auf.

Wir erhalten somit den dreidimensionalen Raum als m\"oglichen ``Lagerort"
gro{\ss}er Anordnungen von bin\"aren Elementen, oder kurz von ``starren
K\"orpern", ganz in Sinne von Poincar\'e (``La science et l'hypoth\`ese")
und Einstein \cite{ae}, der sich hier an Poincar\'e anlehnt.

Bemerkenswert ist, da{\ss} wir Raum-Zeit nicht als eigenst\"andiges
Ph\"anomen, sondern als eine Eigenschaft makroskopischer K\"orper erhalten.
Das hei{\ss}t: bei Abwesenheit von K\"orpern ist Raum-Zeit nicht definiert.

Der Minkowski-Raum ist eine Approximation mittels Gruppenkontraktion,
die f\"ur die Umgebung eines Punktes in Raum-Zeit g\"ultig ist.
Da{\ss} dies nur eine N\"aherung ist, hat Konsequenzen, auf die ich noch
eingehen werde.

Was die quantenmechanische Seite betrifft, so folgen aus der Definition
(\ref{4-4}) der lokalisierten Zust\"ande die \"ublichen
Vertauschungsrelationen zwischen Ort und Impuls, man erh\"alt
Welleneigenschaften und nat\"urlich auch die Heisenbergsche
Unbestimmtheitsrelation.

\section{Teilchen in Raum-Zeit}

Einzelne bin\"are Elemente bzw. Spinoren besitzen, au{\ss}er dem
Spin-Freiheitsgrad, keine weiteren Freiheitsgrade, die wir mit Raum-Zeit
in Verbindung bringen k\"onnen.
\textit{Relativ zu Makrosystemen} lassen sich ihnen aber durchaus
Raum-Zeit-Eigenschaften zuordnen.
Nehmen wir an, wir haben ein Makrosystem, bestehend aus $N$ Spinoren mit
einem Gesamtimpuls $P$, und f\"ugen einen weiteren Spinor hinzu.
Im Rahmen der Poincar\'e-Approximation ergibt das wieder einen makroskopischen
Zustand bestehend aus $N+1$ Spinoren mit einem geringf\"ugig ge\"anderten
Impuls $P + p_s$.
Wir k\"onnen dann sagen: relativ zum Makrosystem wird der Spinor durch
einen Impuls $p_s$ beschrieben.
$p_s$ wird dabei noch von der Spin-Quantenzahl $s$ des Spinors abh\"angen.
Dieses Spiel l\"a{\ss}t sich mit beliebigen anderen Makrosystemen
wiederholten. Dadurch erscheint der Impuls $p_s$ wie eine Eigenschaft des
Spinors selbst.
Wir k\"onnen somit hilfsweise einen neuen Hilbertraum einf\"uhren,
bestehend aus den Impulseigenzust\"anden zu $p_s$, die wir dann formal dem
Spinor zurechnen.
Wir sollten aber nicht vergessen, da{\ss} diese Impulse keine Eigenschaft
des Spinors selbst, sondern eine Eigenschaft der Beziehung des Spinors zum
Makrosystem beinhaltet.

Weil diese Impulse als Impulsdifferenzen eines Makrosystems definiert sind,
\"ubernehmen sie die Transformationseigenschaften des Makrosystems bez\"uglich
der Poincar\'e-Gruppe.
Folglich k\"onnen wir auch mit diesen Zust\"anden lokalisierte Zust\"ande
oder auch Wellenpakete bilden und diese in Raum-Zeit formal auf die Reise
schicken.
Treffen sie dabei auf ein anderes Makrosystem, k\"onnen sie mit diesem
wieder einen gemeinsamen Zustand bilden und damit wird aus dem formalen
Zustand wieder eine reale Impulsdifferenz - dieses Mal relativ zu dem anderen
Makrosystem.

Wir k\"onnen den hier geschilderten Gedanken weiter konkretisieren, indem
wir \"uberlegen, wie diese Impulsdifferenzen mit dem Spinorfreiheitsgrad
zusammenh\"angen.
Makrosystem und einzelner Spinor werden im Rahmen der exakten
SO(3,2)-Symmetrie durch SO(3,2)-Operatoren beschrieben, die auf Zust\"ande
$|S\rangle$ bzw. $|s\rangle$ wirken. Also z.B.
\begin{equation}
(S_{\mu4} + s_{\mu4}) |S\rangle |s\rangle   .               \label{5-1}
\end{equation}
$S_{\mu4}$ wird im Rahmen der Poincar\'e-N\"aherung durch $P_\mu$ ersetzt,
also
\begin{equation}
(P_\mu + s_{\mu4}) |P\rangle |s\rangle   .                  \label{5-2}
\end{equation}
Auf Grund der eben geschilderten \"Uberlegung gilt im Rahmen der
Poincar\'e-N\"aherung auch eine Darstellung in der Form
\begin{equation}
(P_\mu + p_\mu) |P\rangle |p_s\rangle   .                   \label{5-3}
\end{equation}
Es mu{\ss} also eine Beziehung geben
\begin{equation}
s_{\mu4} |s\rangle \Longleftrightarrow p_\mu |p_s\rangle ,  \label{5-4}
\end{equation}
mit der sich der Impuls $p_s$ aus dem Spin $s$ bestimmen lassen sollte.
Diese Beziehung mu{\ss} kovariant bez\"uglich der Poincar\'e-Gruppe sein
und sie sollte linear sowohl hinsichtlich $p_s$ als auch $s$ sein.
Wir suchen folglich nach einer invarianten Operatorbeziehung, die aus
den beiden charakteristischen Operatoren $p_\mu$ und $s_{\mu4}$ zu bilden
ist, in diesen beiden jeweils linear ist und auf einen Zustand $|p_s\rangle$
anzuwenden ist.

Die einzige Invariante, die diesen Anspr\"uchen gen\"ugt, ist
$\gamma^\mu p_\mu$. Angewandt auf den Zustand $|p_s\rangle$ erhalten wir
\begin{equation}
(\gamma^\mu p_\mu - m) |p_s\rangle = 0                      \label{5-5}
\end{equation}
mit einer Konstanten $m$.
Dies ist die Dirac-Gleichung mit der ``Masse" $m$.

Der Wert von $m$ kann folgenderma{\ss}en abgesch\"atzt werden.
Aus der Dirac-Gleichung folgt bekannt\-lich, da{\ss}
\begin{equation}
p^\mu p_\mu = m^2  .                                         \label{5-6}
\end{equation}
Andererseits steht der Operator $p^\mu p_\mu$ f\"ur den Spin-Operator
$\frac{1}{4}\gamma^\mu \gamma_\mu$.
Dieser Operator hat den festen Wert $1$.
Somit ergibt sich der Wert $1$ als Absch\"atzung f\"ur die Masse $m$, woraus
wir zumindest schlie{\ss}en k\"onnen, da{\ss} die Masse endlich ist.
Der Wert ist zu verstehen in Einheiten der Spin-Quantenzahlen. Einen Wert
in der \"ublichen Einheit einer Masse erhalten wir dann, wenn wir uns auf
eine Einheit festlegen, in der wir den Impuls $p$ messen wollen.

Wir sind damit, ausgehend von bin\"aren Elementen, \"uber Spinoren zu
massiven Spin-1/2-Teilchen in Raum-Zeit gekommen.

Bleibt noch etwas zur Statistik zu sagen.
Wir wissen bereits, da{\ss} die Spinoren au{\ss}er durch den
Spin-Freiheitsgrad nicht unterscheidbar sind.
Ein Zustand aus zwei Spinoren darf sich daher, bei der Vertauschung von
zwei Spinoren, bis auf das Vorzeichen nicht \"andern.
Die hinzugekommene raum-zeitliche Beschreibung \"andert daran nichts.
Die Vertauschung von zwei r\"aumlich getrennten Spinoren l\"a{\ss}t sich
aber durch eine Drehung um $180$ Grad um eine Achse senkrecht zu ihrer
Verbindungslinie beschreiben. Mit dem eingef\"uhrten Formalismus kann man
dann leicht nachrechnen, da{\ss} bei einer solchen Drehung jeder
Spinor-Zustand einen Phasenfaktor $i$ erh\"alt. Insgesamt ergibt sich damit
ein Faktor $-1$.
Damit gen\"ugen die Teilchen der Fermi-Statistik.
(Diese Argumentation zur Ableitung des Pauli-Prinzips wurde erstmals von
A. A. Broyles \cite{aab} im Jahr 1976 benutzt und wurde sp\"ater, offenbar
unabh\"angig, auch von Feynman und Weinberg \cite{fw} verwendet.)

\section{Wechselwirkung}

Die Approximation der de~Sitter-Gruppe durch die Poincar\'e-Gruppe hat uns
zu freien Teilchen mit den Eigenschaften von Elektronen und Positronen
gef\"uhrt. Die \"Ubereinstimmung w\"are perfekt, wenn wir auch noch Aussagen
\"uber eine Ladung machen k\"onnten.
Das f\"uhrt uns auf den Begriff der Wechselwirkung.

Um uns diesem Begriff zu n\"ahern, m\"ussen wir uns noch einmal kritisch
mit der G\"ultigkeit der Approximation durch die Poincar\'e-Gruppe befassen.
Bei genauerer Betrachtung stellt man fest, da{\ss} es darum keineswegs
besonders gut bestellt ist.
Die Sache ist zwar in Ordnung f\"ur Makrosysteme; aber ich habe ja nun
auch den einzelnen Spinoren einen Impuls zugeordnet unter Berufung auf
die G\"ultigkeit der Approximation f\"ur Makrosysteme.
Nun transformiert sich der Spin-Freiheitsgrad des Spinors nach wie vor
mit einem Spin-Operator $\frac{1}{2}\gamma_\mu$, und der zugeordnete
Relativimpuls mit $p_\mu$.
Das hei{\ss}t, der vollst\"andige ``Translationsoperator" des Spinors
enth\"alt neben dem Operator $p_\mu$ noch einen $\gamma_\mu$-Term.
In der Literatur \"uber Gruppen\-kontraktion wird angenommen, da{\ss}
letzterer klein gegen\"uber dem Impuls-Operator ist und daher im
Grenz\"ubergang der Kontraktion zu vernachl\"assigen ist.
In unserem Fall beschreibt aber der Impuls $p_s$ gerade den Effekt auf
das Makrosystem, der durch das Hinzuf\"ugen eines Spinors verursacht wird.
Folglich haben die Erwartungswerte der Operatoren $p_\mu$ und
$\frac{1}{2}\gamma_\mu$ die gleiche Gr\"o{\ss}enordnung.
Somit erscheint die Poincar\'e-N\"aherung in diesem Fall mehr als
fragw\"urdig.

Nun, was tut ein theoretischer Physiker, wenn er eine einfache
N\"aherung gefunden hat, dann aber feststellen mu{\ss}, da{\ss} die
wahre Situation doch komplizierter ist?
Er beh\"alt die N\"aherung als Basistheorie bei und betrachtet die Terme,
die seine N\"aherung st\"oren, als ``kleine" St\"orung.
In unserem Fall besteht die St\"orung aus dem Operator $\frac{1}{2}\gamma_\mu$,
der neben dem gewohnten Translations\-operator $p_\mu$ auftaucht.
Es l\"a{\ss}t sich nun zeigen\cite{ws}, da{\ss} eine st\"orungstheoretische
Behandlung des $\gamma_\mu$-Terms dazu f\"uhrt, da{\ss} die Teilchen ihre
Freiheit verlieren und miteinander in Wechselwirkung treten.
Diese Wechselwirkung ist empirisch und theoretisch wohlbekannt.
Es ist dies die elektromagnetische Wechselwirkung, und die St\"orungtheorie
f\"ur den $\gamma_\mu$-Term ist formal identisch mit der
Quantenelektrodynamik.

Die St\"arke dieser Wechselwirkung ist im Gegensatz zum Standardmodell durch
die Theorie be\-stimmt und l\"a{\ss}t sich absch\"atzen. Eine solche
Absch\"atzung liefert f\"ur die Kopplungskonstante einen Wert, der mit dem
experimentellen Wert der Sommerfeldschen Feinstrukturkonstante $\alpha$ bis
auf sieben Dezimalstellen \"ubereinstimmt.
Tats\"achlich reproduziert diese Absch\"atzung Wylers\cite{aw}
semi-empirische Formel f\"ur $\alpha$ aus dem Jahr 1969, ohne jedoch an
Wylers (fehlgeschlagenen) Versuch einer Begr\"undung anzukn\"upfen.
Damit k\"onnen wir den Teilchen nun auch eine Ladung und zwar in der
Gr\"o{\ss}e der elektrischen Elementarladung zuordnen.

Wir k\"onnen somit die eingangs gestellt Frage ``Kann man die Bits der
Teilchenphysik beobachten?" positiv beantworten: Diese Bits zeigen,
wenn wir sie gem\"a{\ss} Bohr relativ zu makroskopischen Anordnungen
beobachten, alle Eigenschaften von Elektronen und Positronen.

Die Ableitung der Quantenelektrodynamik in \cite{ws} demonstriert, da{\ss}
der informationstheoretische Zugang zur Teilchenphysik m\"achtig genug ist,
nicht nur die Quantenmechanik, sondern auch eine realistische
Quantenfeldtheorie hervorzubringen.

Bei eingehenderer Betrachtung lassen sich noch weitere St\"orungsterme
identifizieren, die auf Grund ihrer Charakteristiken den verbleibenden
Wechselwirkungen zuzuordnen sind.
Bemerkenswerterweise schlie{\ss}t dies, neben schwacher und starker
Wechselwirkung, auch die Gravitation ein, f\"ur die man klare Vorgaben
hinsichtlich der Formulierung einer Quantengravitation erh\"alt.

N\"aheres finden Sie in meinem Beitrag \cite{ws} zum Band
``Progress in General Relativity and Quantum Cosmology Research",
der 2005 bei Nova Science Publishers, New York, erscheinen soll.

\section{Resume\'e}

Es ist in der Vergangenheit verschiedentlich die Ansicht ge\"au{\ss}ert
worden, da{\ss} eine fundamentale Theorie es erlauben sollte, zugleich
Quantenmechanik und Raum-Zeit aus grundlegenderen Prinzi\-pien abzuleiten.

Ich habe Ihnen heute, mit Bohrs Hilfe, solche Prinzipien vorgestellt.
Es sind dies:
\begin{enumerate}
\item
Die Physik besteht aus der Gewinnung und logischen Verkn\"upfung
von \begin{em}Information\end{em} \"uber die Natur.
\item
Der Gegenstand der Teilchenphysik sind die \begin{em}kleinsten\end{em}
Bausteine dieser Information.
\item
Die Informationsgewinnung geschieht \begin{em}relativ\end{em} zu
makroskopischen Anordnungen.
\end{enumerate}

Diese Prinzipien stellen keine abstrakten Axiome dar, sondern
beschreiben konkret, was wir tun, wenn wir Elementarteilchenphysik
betreiben.
Sie lassen sich somit nachpr\"ufen, und ich denke, sie geben die Grundlagen
unserer Disziplin richtig wieder.

Das erste Prinzip definiert das Gebiet, auf dem wir uns bewegen: Es geht um
Information \"uber die Natur.
Wir sammeln Informationen, modellieren Zusammenh\"ange und leiten weitere
Informationen daraus ab.

Das zweite Prinzip f\"uhrt auf die logischen Grundbausteine der Information,
die bin\"aren Elemente oder Bits, und legt dabei, wie wir gesehen haben, die
Basis f\"ur grundlegende Symmetrieeigenschaften.

Das dritte Prinzip liefert uns die quantenmechanische Beschreibung
als Werkzeug zur Behandlung ``relativer" Information.
Diese Art der Beschreibung f\"uhrt schlie{\ss}lich \"uber die erw\"ahnten
Symmetrieeigenschaften in erster N\"aherung zu einem relativistischen
Raum-Zeit-Kontinuum und zu Teilchen in Raum-Zeit.
In h\"oherer N\"aherung erh\"alt man dann vier fundamentale Wechselwirkungen.

Bei diesen Schlu{\ss}folgerungen mu{\ss}ten wir kein ``Naturgesetz"
bem\"uhen.
Eingegangen ist lediglich, da{\ss} beide Orientierungen eines bin\"aren
Elements, sowie beide Kodierungsarten von bin\"arer Information
gleich\-berechtigt zu behandeln sind.
Eingegangen ist weiter die Entscheidung f\"ur einen Hilbertraum-Formalismus
zur Darstellung von Mischzust\"anden.
Dar\"uberhinaus sind die einzigen Gesetze, die eingehen, die der
mathematischen Logik und diese sind klar und begreifbar.

Mit der Entscheidung f\"ur einen Hilbertraum-Formalismus haben wir uns
ein Begriffssystem verschafft, das es erlaubt, uns in der gewohnten
Sprache der Physik auszudr\"ucken, anstatt mit Bits zu hantieren.
Seine Einf\"uhrung bedeutet keine zus\"atzliche Annahme, sondern
dient lediglich der klaren Formulierung von Beziehungen zwischen Mengen
von bin\"aren Elementen.
Der Sinn dieser Entscheidung mi{\ss}t sich daran, wie weit sie uns auf
Strukturen f\"uhrt, die wir mit solchen der Empirie vergleichen k\"onnen.
Die Entscheidung f\"ur einen Hilbertraum ist also pragmatischer Natur
und nur a-posteriori zu rechtfertigen.
In der Tat enth\"alt diese Entscheidung Sinn dadurch, da{\ss} sie uns auf ein
Raum-Zeit-Kontinuum f\"uhrt und wir in dieser Raum-Zeit eine Wechselwirkung
finden, die der elektromagneti\-schen entspricht. Diese Wechselwirkung
wiederum erlaubt es, das Raum-Zeit-Kontinuum auszumessen und ihm auf diese
Weise physikalische Realit\"at zu verleihen.

Damit ergibt sich das folgende Bild: sobald die drei Prinzipien f\"ur richtig
befunden sind, und bei den Schlu{\ss}folgerungen daraus keine logischen
Fehler unterlaufen, sind die geschilderten Folgerungen weitestgehend
\begin{em}zwingend\end{em}.

Das hei{\ss}t: Die physikalische Wirklichkeit, wie sie sich einem
Teilchen-Physiker darstellt, ergibt sich nicht als Ausdruck eines
fundamentalen Naturgesetzes bzw. einer ``Weltformel", sondern als
zwangsl\"aufige, logische Konsequenz unserer Art Teilchenphysik zu betreiben.

Das impliziert dann auch, da{\ss} diese Grundprinzipien Bestandteil einer
jeden realistischen Ele\-mentarteilchentheorie sein m\"ussen.
Ob in einer zuk\"unftigen, vollst\"andigen Theorie noch weitere Prinzipien
zum Tragen kommen oder zus\"atzliche Annahmen erforderlich sein werden, ist
derzeit offen.
Allerdings habe ich vorl\"aufig keinen Hinweis auf die Notwendigkeit
konzeptioneller Erg\"anzungen.

Um noch einmal auf die verwendeten logischen Grundbausteine zur\"uckzukommen:
Sie bestehen aus der einfachsten und allgemeinsten Struktur, in die sich
Information kodieren l\"a{\ss}t.
Da diese Grundbausteine (bei Betrachtung relativ zu Makrosystemen, wie
oben beschrieben) alle Eigenschaften von Elektronen aufweisen, m\"ussen wir
folgern, da{\ss} Elektronen (vermutlich zusammen mit Quarks) die elementarsten
physikalischen Strukturen darstellen.
Dies impliziert, da{\ss} es keine noch elementareren Strukturen geben kann,
\"uber die wir durch ein physikalisches Experiment Informationen erhalten
k\"onnen.
Das hei{\ss}t, der Weg zu immer elementare\-ren Gebilden, den die Physik in
den letzten hundert Jahren verfolgt hat, findet hier ein nat\"urliches Ende.

Damit wird insbesondere der Spekulation \"uber elementare Strukturen
im Bereich der Planck-L\"ange weitgehend der Boden entzogen.
Genauer gesagt: Falls solchen Strukturen eine, wie auch immer geartete, Form
der Existenz zukommen sollte, best\"unde trotzdem keine logische Verbindung
zu dem, was wir im Sinne von Bohr unter Physik verstehen.
Damit w\"aren, zum einen, solche Strukturen der physikalischen Erfahrung
grunds\"atzlich nicht zug\"anglich.
Zum anderen w\"are ihre Einbeziehung in eine physikalische Theorie unn\"otig
und w\"urde dem Prinzip der Denk\"okonomie widersprechen.

Dieser Beitrag hat eine Grenze unserer Erfahrungsm\"oglichkeiten aufgezeigt.
Aber diese Grenze stellt keine Einschr\"ankung in unserem Streben nach
Erkenntnis dar.
Sie bedeutet lediglich, da{\ss} es jenseits dieser Grenze nichts gibt, was
wir in Erfahrung bringen k\"onnen.
Tats\"achlich finden wir an dieser Grenze eine klare und wohlverstandene
Basisstruktur vor, die uns die Chance er\"offnet, darauf aufbauend eine
durchg\"angige, konsistente Elementarteilchentheorie zu formulieren, die
dann auch die Gravitation umfa{\ss}t.

Ich schlie{\ss}e mit einem weiteren Zitat von Niels Bohr:
\begin{quote}
``Es ist die Aufgabe der theoretischen Wissenschaften, tiefe Wahrheiten
zu Trivialit\"aten zu reduzieren."
\end{quote}
Ich hoffe, hierzu ein wenig beigetragen zu haben. Siehe auch \cite{ws1}.

\renewcommand{\baselinestretch}{1.1}

\end{document}